\documentclass[12pt]{amsart}
\def\acts{\triangleright}
\def\Z{{\mathbb Z}}
\def\C{{\mathbb C}}
\def\R{{\mathbb R}}
\def\CA{{\mathcal A}}

\def\CJ{{\mathcal J}}

\def\CS{{\mathcal S}}

\DeclareMathSymbol\crossrt{\mathrel}{AMSb}{"6E}
\DeclareMathSymbol\crosslt{\mathrel}{AMSb}{"6F}

\def\ts{\otimes}

\def\oh{\hbox{$\frac{1}{2}$}}

\def\const{\hbox{const}}

\newtheorem{lemma}{Lemma}[section]

\newtheorem{proposition}[lemma]{Proposition}
\newtheorem{corollary}[lemma]{Corollary}
\theoremstyle{definition}
\newtheorem{example}[lemma]{Example}

\newtheorem{observation}[lemma]{Observation}

\usepackage{amscd}
\title{Dynamical Noncommutative Spheres}
\author{Andrzej Sitarz}
\email{Andrzej.Sitarz@th.u-psud.fr}
\address{
\begin{itemize}
\item[${}^{a}$] Laboratoire de Physique Theorique, Universit\'e 
Paris-Sud, Bat. 210, 91405  ORSAY Cedex, France 
\item[${}^{b}$] Institute of Physics, Jagiellonian University,  
Reymonta 4, 30059 Krak\'ow, Poland
 \end{itemize}}
 
\thanks{${}^\dagger$\ Supported by Marie Curie Fellowship.}
\begin{document}
\begin{abstract}
We introduce a family of noncommutative $4$-spheres, 
such that the instanton projector has its first Chern class 
trivial: $ch_1(e) = B \chi + b \xi$. We construct for them 
a 4-dimensional cycle and calculate explicitly the Chern-Connes 
paring for the instanton projector.
\end{abstract}

\maketitle

\section{Introduction}

The construction of noncommutative spheres based on homological
principles was proposed by Connes \cite{Connes2000}, the basic 
assumption is that the algebra is generated by the elements of 
an projector (or unitary matrix over the algebra in the odd case) and 
its Chern classes in Hochschild homology vanish in all dimensions  
smaller than the dimension of the manifold.

Connes proved that in dimension $2$ only commutative solutions appear. 
First noncommutative examples of solutions in dimension three and four 
were constructed in \cite{CoLa} then a systematic analysis of this type
of solutions as well as construction of all three-dimensional solutions 
were given in \cite{CoDuVio}. All constructed examples of noncommutative 
three (and four)  are of good homological dimension (related to Hochschild 
or cyclic homology).  Moreover, they seem to be (and in some cases 
certainly are) nice examples of noncommutative spin geometries, as 
defined by Connes \cite{Connes2}.

In this paper we introduce a variation of the noncommutative deformation 
of a four-sphere. With a subtle generalization of the deformation parameter
we shall obtain a family of objects indexed by smooth functions on an 
interval, a special case of a constant function corresponding to the 
isospectral deformation \cite{CoLa}. The deformation in question 
goes beyond the so far considered models fo noncommutative 
spheres like $SU_q(2)$ and its suspension (see \cite{DLM}), 
deformations based on suspensions (and their twists) of Podles spheres 
(\cite{Si0,BG}) or the above mentioned isospectral deformations. It 
rather extends the original ideas of Matsumoto \cite{Matsumoto} 
who first considered (in $C^*$-algebraic setup) 
the three-spheres\footnote{The original definition of Matsumoto 
three-spheres uses different generators, 
however, in $C^*$-algebraic formulations invertible 
transformations between generators of \cite{Matsumoto} and 
\cite{CoLa} could be easily constructed explicitly: these 
are however only continuous but not smooth.},
studied later in \cite{CoLa}; in fact he described an entire 
family of their generalizations (we shall mention them later).

In the paper we present the construction of the deformation, define
the instanton projector, differential calculus over the deformed 
spheres, we construct a four-dimensional cycle,
calculate Chern classes of the instanton projector and the 
corresponding Chern numbers.

The name {\em dynamical}, which we use for the deformation 
has been motivated by possible physical applications: although
we work here with a deformation of a compact manifold, it is easy
to generalize the procedure to construct such deformations of
$\R^4$ or $M \times \R$. With the natural interpretation of the 
coordinate as time we obtain {\em time-dependent 
noncommutativity}, an idea, which could be motivated, for instance, 
in string theory from considerations of branes in a non-static $B$-field.

\section{Preliminaries}

We shall begin by recalling the main steps of the construction 
of  isospectral deformations, as done in \cite{CoLa}. Let 
$M$ be a compact manifold, $\CA = C^\infty(M)$ and 
let the two torus $T^2$ act on $\CA$. 

Since any smooth function (with respect to the action 
of the torus) could be presented as a doubly infinite 
norm convergent series of homogeneous elements, where
$f$ is homogeneous of of bidegree $n_1,n_2$ iff
$$ (u_1, u_2) \acts f = (u_1)^{n_1} (u_2)^{n_2} f, $$
for $u_1,u_2 \in T^2$, one might introduce a deformed
algebra as using a left (or right) twist maps:

\begin{equation}
\sum_{n_1, n_2} T_{n_1,n_2}  = T \mapsto l(T) =  \sum_{n_1, n_2} 
T_{n_1,n_2} \lambda^{n_2 \delta_1} 
\end{equation}
where $\lambda$ is a complex number of module $1$ and 
$\delta_1, \delta_2$ are the generators representations of 
the (projective) unitary representation of the action of the
torus.

Then we have the lemma:
\begin{lemma}[\cite{CoLa}, Lemma 4]
There exists an associative product on the vector space of
smooth functions $\CA$, $(x,y) \mapsto x*y$ such that
\begin{equation}
l(x) l(y) = l(x*y).
\end{equation}
For the homogeneous elements $(x,y)$ of order $n_1, n_2$
and $m_1,m_2$, respectively, it is:
\begin{equation}
x*y = \lambda^{n_1 m_2} xy.
\end{equation}
\end{lemma}

From the algebraic point of view the constructed deformation is 
a cocycle deformation of the algebra through the twist from the Cartan 
subalgebra of its symmetry group. This description was developed 
in \cite{Sit} and used to demonstrate that twisted isometry of the 
algebra is the Hopf-algebra isometry of deformed spectral triple,
a dual approach to symmetries was suggested in \cite{Varilly}, 
whereas a systematic approach to $\theta$-deformations is presented
in \cite{CoDuVio}.

\subsection{Dynamical twists}

We shall introduce here a generalization of the above deformation, 
which we shall study in details in a particular case of the sphere. 
Our assumptions are as in the situation discussed earlier: we work 
with smooth functions on a compact oriented manifold, such that 
its symmetry groups contains a torus and we assume that the 
(smooth) action of the isometry group is (projectively) lifted to 
the Hilbert space.

Let $f$ be a smooth function, and let $T(f)_{n_1,n_2}$ be 
its component of the Fourier series with respect to the action of $T^2$:
$$ T(f) = \sum_{n_1,n_2} T(f)_{n_1,n_2},$$
where the series is norm converging (norms of homogeneous 
elements, which are the elements of this series, are of rapid decay).

Let $H$ be a self-adjoint element of the algebra $C^\infty(M)$
(real smooth function), which is of bidegree $(0,0)$, so it is invariant 
with respect to the action of the two-torus.

Let us define a map, which shall assign to every element 
$f \in C^\infty(M)$ an element of the deformed algebra:

\begin{equation}
T_H(f) = \sum_{n_1,n_2} T(f)_{n_1,n_2} e^{2\pi i n_2 H \delta_1}.
\end{equation}

Let us observe that the series is again infinite norm 
convergent  (we modify each element by multiplication
with an operator of norm $1$) and since $H$ commutes with 
the action of torus the definition well posed (that is the
bidegree of an element is stable under multiplication
by any function of $H$). So we have a lemma:

\begin{lemma}\label{dytwi}
If $f,g$ are homogeneous operators of degrees $(n_1,n_2)$
and $(m_1,m_2)$, respectively, then:
\begin{equation}
T_H(f) T_H(g) = T_H(f \ast  g),
\end{equation}
where 
\begin{equation}
f \ast g = e^{2\pi i H n_2 m_1} (f g).
\end{equation}

Similarly, one may define an opposite deformation:
\begin{equation}
T_H^o(f) = \sum_{n_1,n_2} e^{2\pi i n_1 H \delta_2} T(f)_{n_1,n_2} .
\end{equation}
such that $[T_H^0(x), T_H(y)]=0$ if $[x,y]=0$
\end{lemma}
\begin{proof}
The proof of the lemma follows directly the proof of lemma 4 
of \cite{CoLa}.
\end{proof}

We shall present now two basic examples of this type of
deformation.

\begin{example}[Three-torus and Heisenberg group algebra]
Let $T^3$ be a three-torus, consider the natural 
action of two-torus $T^2 \subset T^3$ on $C^\infty(T^3)$.
If we denote the unitary generators of $C^\infty(T^3)$ by 
$U,V,W$, then $W$ remains the invariant element under
the action of $T^2$.

If we make the choice of $H=\theta$ as a constant we obtain a
product of a noncommutative torus $T^2_\theta$ with $S^1$. 
However, the simplest nontrivial choice of $e^{2\pi i H} = W$
gives us the algebra relations:
\begin{equation}
UV = W VU, \;\;\;\; [U,W] = [V,W] = 0. \label{Heis}
\end{equation}

Clearly, the first relation can be generalized  to: 
\begin{equation}
UV = f(W) VU. 
\end{equation}
where $f(W)$ is a suitable smooth function 
$f: S^1 \to S^1$, however, in the particular case 
$f(W)=W$ the algebra is the group algebra of the 
discrete 3-dimensional Heisenberg group 
\cite{AnPa}. We shall study the properties of this
algebra, in particular the explicit construction of
the $K$-cycle and Chern-Connes pairing in a 
separate paper \cite{Sit2}
\end{example}

\begin{example}[The 4-sphere]
Let us consider a $\ast$-algebra generated 
by elements $a,a^\ast,b,b^\ast$ 
and $t=t^\ast$ subject to the following 
set of relations:

\begin{equation}
\begin{array}{lll}
[a, t] =0,  & & [a^\ast, t] = 0, \\
{}[b, t] =0,  & & [b^\ast, t] = 0, \\
{}[a ,a^\ast] =0, & \phantom{xxxxx} & [b, b^\ast] =0, \\
ab = \lambda(t) ba, &  & a b^\ast = \bar{\lambda}(t) b^\ast a, \\
a^\ast b = \bar{\lambda}(t) a^\ast b, & & a^\ast b^\ast =
\lambda(t) b^\ast a^\ast,
\end{array} \label{algebra}
\end{equation}
where $\lambda(t)$ is a unitary element, 
$\lambda(t) \bar{\lambda}(t) =1$, expressed as a function 
of the central element $t$, so we may assume:
\begin{equation}
\lambda(t) = e^{-i\phi(t)},
\end{equation}
where $\phi$ is a smooth real function of $-1 \geq \leq 1$.

Furthermore, we have the restriction:
\begin{equation}
a a^\ast + b b^\ast + t^2 = 1, \label{sphere}
\end{equation}
which is the relation defining the (noncommutative) 4-sphere.

One could easily verify that the above set of relations is consistent, 
for any choice of the function $\phi$, the particular example of 
$\phi=\theta=\const$ being the isospectral deformation of the 
sphere. 

Passing from algebraic (polynomial) algebra to the algebra of smooth
functions one can easily observe that the algebra describes the 
dynamical deformation of the four-sphere as presented in 
Lemma \ref{dytwi}, with $H = \phi(t)$ (the parameter $t$ corresponds 
to the choice of presentation of $S^4$ as a suspension of $S^3$).

We shall denote this algebra by $\CS^4_\lambda$, let us observe that 
the center of the  algebra in question contains $t$, $a a^\ast$ and 
$b b^\ast$ but could be much bigger depending on the function 
$\lambda$. 
\end{example}

\section{Instanton bundles over $\CS^4_\lambda$}

One of the most appealing feature of the construction 
of \cite{CoLa} was the existence of the instanton bundle 
over the deformed algebra. This was shown by the construction 
of the projector $e$ with vanishing lower Chern classes 
and $ch_2(e)$ giving rise to a Hochschild cocycle over 
the algebra.

The projector in our case is unmodified:

\begin{equation}
e = \frac{1}{2} \left( \begin{array}{llll}
1+t & 0 & a & b \\
0 & 1+t & - \lambda(t) b^\ast & a^\ast \\
a^\ast & - \bar{\lambda}(t) b & 1-t & 0 \\
b^\ast & a & 0 & 1-t \end{array} \right), \label{proj}
\end{equation}

the only significant distinction for the $\lambda=\const$ case
is that no longer all the entries of the projector are the generators
of the algebra. Of course, since $\lambda$ is not a constant 
parameter one may easily verify that the Chern homology 
elements constructed out of $e$ shall not be the same as in 
$\lambda=\const$ case. In particular, we have:

\begin{equation}
ch_1(e) = t \ts x_i \ts y_i - x_i \ts t \ts y_i + x_i \ts y_i \ts t,
\end{equation}
where:
\begin{equation}
x_i \ts y_i = b \ts b^* - b^* \ts b +
\lambda b^* \ts \overline{\lambda} b  -  
\overline{\lambda} b \ts \lambda b^*.
\end{equation}
 
It is easy to verify that $b\, ch_1(e)$ vanishes, however
$B\, ch_1(e)$ does not:
\begin{equation}
B \, ch_1(e) = 1 \ts ch_1(e). 
\end{equation}

We shall postpone further discussion of the Chern classes 
until the last section of the paper, when it shall be clear that
although $ch_1(e)$ does not vanish, its class is trivial.

In fact, using the  the natural construction of differential 
structures on the deformed  sphere and the natural trace 
on the algebra we shall give explicit formula for the volume 
form, which arises naturally from the Chern 
class $e\,de\,de\,de\,de$ and calculate the Chern number 
of the above projector $e$.

\section{The differential calculus on $S^4_\lambda$}

Unlike in the $\lambda=\const$ case we have no clear 
indication for the construction of differential calculi. 
We shall look for a guiding principle of the smallest
calculi, which, when restricted to commutative subalgebras, 
remains classical and for $\lambda=1$ gives the correct
limit of the differential structures on a four-sphere.

Before we begin let us observe that the commutation relations 
between algebra generators $a,a^*,b,b^*$ could be rewritten as
\begin{equation}
x^i x^j = A_{ij} \, x^j x^i, \;\;\; 1 \leq i,j \leq 4, \label{comm}
\end{equation}
where there is no summation in the formula, $x^i$ denote
the generating monomials $a,a^*,b,b^*$ and the matrix 
$A_{ij}$ is $t$-dependent,  in our case:
\begin{equation}
A = \left( \begin{array}{llll}
1 & 1 & \lambda(t) & \bar{\lambda}(t) \\
 1 & 1 & \bar{\lambda}(t) &\lambda(t)  \\
 \bar{\lambda}(t) &\lambda(t) &  1 & 1  \\
\lambda(t) &   \bar{\lambda}(t) &1 & 1 
\end{array} \right).
\end{equation}

We make an Ansatz that the bimodule of one forms 
is generated by $dx^i$ and a central one-form $dt$, with
quadratic the bimodule commutation rules: 
\begin{equation}
x^i \, dx^j = A_{ij} dx^j \, x^i + \oh B_{ij} dt (x^i \, x^j). \label {diff1}
\end{equation}

We assume as well, that $t\, dx^i = dx^i\, t$. It is easy to see that 
such relations are consistent with the algebra commutation rules.
Further, if we differentiate (\ref{comm}) and use (\ref{diff1}) we 
obtain the following relation between $B$ and $A$:
\begin{equation}
\oh ( B_{ij} - B_{ji} ) = \frac{1}{A_{ij}} \dot{A}_{ij},
\end{equation} 

We shall restrict ourselves only to the antisymmetric solution
for $B$, which are explicitly given by the above formula.

Expressing the relations (\ref{diff1}) in terms of the generators we have:
\begin{equation}
\begin{array}{lll}
a\, da = da\, a, &\hbox to 1cm{\hfil}& b\, db = db\, b, \\
a\, da^\ast = da^\ast a, & &b\, db^\ast = db^\ast\, b, 
\end{array}
\end{equation}
and
\begin{equation}
\begin{array}{l}
a\, db = \lambda(t)\, db\, a + \oh \dot{\lambda}(t) \bar{\lambda}(t)\, dt\, ab, \\
a\, db^* = \bar{\lambda}(t)\, db^*\, a - \oh \dot{\lambda}(t) \bar{\lambda}(t)\, dt\, ab^* \\
b\, da = \bar{\lambda}(t)\, da\, b - \oh \dot{\lambda}(t) \bar{\lambda}(t)\, dt\, ba, \\
b \,da^* = \lambda(t) \, da^* \, b + \oh \dot{\lambda}(t) \bar{\lambda}(t) \, dt \, ba^* \\
a^\ast\, db = \bar{\lambda}(t)\, db\, a^\ast  - \oh \dot{\lambda}(t)\bar{\lambda}(t)\, dt\, a^\ast b, \\
b^\ast\, da  = \lambda(t)\,  da\, b^\ast + \oh \dot{\lambda}(t)\bar{\lambda}(t)\, dt\, ab^\ast, \\
a^\ast\, db^\ast = \lambda(t)\, db^\ast\, a^\ast  + \oh \dot{\lambda}(t)\bar{\lambda}(t)\, dt\, a^\ast b^\ast, \\
b^\ast\, da^\ast  = \bar{\lambda(t)}\,  da^\ast\, b^\ast - \oh \dot{\lambda}(t)\bar{\lambda}(t)\, dt\, a^\ast b^\ast.
\end{array} \label{diff-e}
\end{equation}

We shall not forget that by differentiating the constraint 
(\ref{sphere}) we have (after using (\ref{diff1})):
\begin{equation}
a\, da^* + a^*\, da + b\, db^* + b^*\, db + 2 t dt = 0, \label{diff2}
\end{equation}

Note that the left-hand side of (\ref{diff2}) side is a central element 
of the bimodule of one forms and therefore the restriction 
(\ref{diff2}) is compatible with the (\ref{diff-e}). Now, we are 
prepared to construct the full differential algebra.

\begin{proposition}
Let $\Omega_u(\CS^4_\lambda)$ be a universal differential algebra, and let 
$\CJ_1 \subset \Omega^1_u(\CS^4_\lambda)$ be the kernel of the
projection map $\pi: \Omega^1_u(\CS^4_\lambda) \mapsto 
\Omega^1(\CS^4_\lambda)$. Then the differential algebra 
$\Omega(\CS^4_\lambda)$ is a $\Z$-graded algebra obtained as
a quotient of $\Omega_u(\CS^4_\lambda)$ by the differential
ideal generated by $\CJ^1 +d \CJ^1$.
\end{proposition}

Clearly, the subbimodule $\CJ^1$  is in our case defined 
by relations (\ref{diff1}) and (\ref{diff2}). Thus by differentiating 
them we obtain the first set of rules:

\begin{eqnarray}
&& dx^i \, dt = - dt \, dx^i, \label{diff3} \\
&& dt\, dt = 0. \\
&& dx^i\, dx^j = - A_{ij} dx^j \, dx^i + \oh B_{ij} A_{ij} dt\, dx^j \, x^i
- \oh B_{ij} dt\, dx^i \, x^j 
\end{eqnarray}
We immediately see that in the differential algebra $\Omega(\CS^4_\lambda)$ 
all generators $dx^i$ and $dt$ are nilpotent, and $da,da^*$, 
$db,db^*$ are pairwise skew-symmetric:
$$ da\, da^* = - da^* \, da, \;\;\;\;\; db\, db^* = - db^* \, db. \label{diff4} $$

For the remaining relations we have:

\begin{equation}
\begin{array}{l}
da\, db + \lambda(t)\, db\, da = \oh \dot{\lambda}(t)\, dt\, db\, a -\oh 
\dot{\lambda}(t) \bar{\lambda}(t) \, dt\,da\, b, \\
db\, da^\ast + \lambda(t)\, da^\ast\, db = \oh \dot{\lambda}(t)\, dt\, da^\ast\, b
- \oh \dot{\lambda}(t)\bar{\lambda}(t)\, dt\, db\,a^\ast.
\end{array}
\end{equation}

Before we prove more results on the differential algebra we 
introduced, let us observe interesting relations:

$$
\begin{array}{l}
b \, da\, da^* = da\, da^*\, b - \oh \dot{\lambda}(t) \bar{\lambda}(t)\, dt\, 
( a\, da^*\, b + a^* \, da\, b) = \\
\phantom{xxxx} = da\, da^*\, b + \oh \dot{\lambda}(t) \bar{\lambda}(t)\, dt \,
( db^*\, b^2 + db\, bb^*).
\end{array}
$$
where in the last step we used (\ref{diff2}).

By differentiating it we obtain:
$$  db \, da\, da^* = da\, da^*\, db  - \oh \dot{\lambda}(t) \bar{\lambda}(t)
\, dt \, db\, db^* b.$$
Similar result can also be proven for $db^*$:
$$  db^* \, da\, da^* = da\, da^*\, db^* 
- \oh \dot{\lambda}(t) \bar{\lambda}(t) \, dt \, db\, db^* b.$$ 
and for products of $da\, db\, db^*$. In particular, we can 
see that:
\begin{equation}
\begin{array}{l}
da\, da^*\, db \, db^* = db\, da\, da^*\, db^* = db^*\, da\, da^*\, db, \\
da\, da^*\, db \, db^* = da\, db\, db^*\, da^* = da^*\, db\, db^*\, da, \\
da\, da^*\, db \, db^* = db \, db^*\, da\, da 
\end{array} \label{3forms}
\end{equation}

Next we shall prove that the differential algebra has a finite 
dimension:
\begin{lemma}
The differential algebra $\Omega(\CS^4_\lambda)$ has dimension $4$,  
for all $n >4$ we have $\Omega^n(\CS^4_\lambda)=0$.
\end{lemma}
\begin{proof}
Clearly, it is sufficient to show that $dt\, da\, da^*\, db\, db^*$ 
vanishes. Let us consider the relation (\ref{diff2}) and multiply
it from the left by a two-form $\oh t da \, da^*$ and from the right 
by $db \, db^*$.

Using the associativity of the product together with relations 
(\ref{diff3}) and the fact that all one generating one-forms 
are nilpotent we obtain:
\begin{equation}
 t^2 dt\, da\, da^* \, db \, db^* = 0. \label{d1}
 \end{equation}
 Similarly, if we multiply (\ref{diff2}) from the left by $dt\, da\, a^*$
and by $db \, db^*$ from the right we obtain:
\begin{equation}
a a^* \, dt\, da\, da^*\, db\, db^* = 0. \label{d2}
\end{equation}
Finally, multiplying it by $b \, dt\, da\, da^*$ from the left and by $db^*$
from the right  we get:
\begin{equation}
b b^* \, dt\, da\, da^*\, db\, db^* = 0. \label{d3}
\end{equation}
By adding the three identities (\ref{d1})-(\ref{d3}) and using
the constraint (\ref{sphere}) we obtain the desired result.
\end{proof}

So far we have shown that the maximal degree of forms is $4$,
it appears however that the structure is exactly as in the "classical"
case and we are able to demonstrate that there exist one 
generating four-form:

\begin{lemma}
The bimodule of differential forms of degree $4$ is a free 
bimodule module over the algebra.
The generating form $\omega$ can be chosen as:
\begin{equation}
\omega = \frac{1}{4} \left( t \, da\, da^* \, db \, db^* 
- 2 a\, dt\, da^*\, db \, db^* + 2 dt \, da\, da^* \, db \, b^* \right),  \label{volf}
\end{equation}
\end{lemma}
where the factor $\frac{1}{4}$ was chosen so that it would correspond
to the volume form on $S^4$ in the classical limit.

\begin{proof}
Consider $t\, \omega$. Using the commutation rules of $dt$ with
other one-forms (\ref{diff3}) as well as the fact that $t$ is central
we might rewrite it conveniently as:
 $$ t \, \omega = \frac{1}{4}( t^2 \, da\, da^* \, db \, db^* +  a \,  da^* \, (2t\, dt) \,db \, db^*
 +  da\, da^* \, (2t dt) \, db \, b^*) = \ldots $$
Next, using (\ref{diff2}) and keeping in mind that $dt$ and $dx^i$ are
nilpotent we get:
 $$ \ldots = \frac{1}{4}(t^2 + a a^* + bb^*) da\, da^* \, db \, db^* = 
 \frac{1}{4} da\, da^* \, db \, db^*.$$
 where we have used first the fact that $a, a^*$ commute with $da,da^*$ 
 (and similar property of $b,b^*$ and their differentials) as well as the
 defining relation (\ref{sphere}).

Similarly one may verify the identities:
\begin{eqnarray}
&&  a \, \omega = \frac{1}{2} \, dt\, da \, db\, db^*, \\
&&  a^* \, \omega = - \frac{1}{2} \, dt\, da^* \, db\, db^*, \\
&& \omega\, b= \frac{1}{2} \, dt\, da \, da^* \, db, \\
&& \omega \, b^* = - \frac{1}{2} \, dt\, da \, da^* \, db^*.
\end{eqnarray}
\end{proof}

The form $\omega$ is central, i.e.~it commutes with all elements
of the algebra. As this result is not evident though it follows 
from an easy algebraic calculation we shall demonstrate
it only for $[b, \omega]$. First, observe that only the first 
component in the sum (\ref{volf}) might give a nontrivial 
contribution as the remaining two contain $dt$ and then 
the nontrivial permutation  rules of generators through 
differentials are homogeneous and will cancel out. 
\begin{equation*}
\begin{array}{l}
[b, \omega] = \frac{1}{4}( b t\, da\, da^* \, db\, db^* - t\,  da\, da^* \, db\, db^* b) = \\
\phantom{xxxx} = \frac{1}{4} t \dot{\lambda}(t) \bar{\lambda}(t) \, 
\left( \bar{\lambda}(t)\, da\, dt \, (ba^*) - dt\, da^* \, (ab) \right) \,db \, db^* = \ldots
\end{array}
\end{equation*}
now, if we permute $t$ and use (\ref{diff2}) to substitute a nontrivial
one-form for $t\, dt$, still using the fact that the one forms are nilpotent:
\begin{equation*}
\begin{array}{l}
\phantom{xxxx} = \ldots \frac{1}{16}  \dot{\lambda}(t) \bar{\lambda}(t) \, 
\left( - \bar{\lambda}(t)\, da\, (a\, da^*) \, (ba^*) +
a^* \, da\, da^* \, (ab) \right) \,db \, db^* = \\ 
\phantom{xxxx} = \frac{1}{16}  \dot{\lambda}(t) \bar{\lambda}(t) \, 
\left( - da\, (a da^*) \,  (a^* b) + a^* \, da \, da^* \, (ab) \right) \,db \, db^* = 0.
\end{array}
\end{equation*}

Before we proceed with the construction of the integral of $4$-forms,
let us observe the properties of a trace on the algebra itself.

\begin{proposition}\label{grtrace}
Let $\int$ be the standard (normalized) integral on $S^4$ 
and $\eta$ be a linear map on $\CS^4_\lambda$, which maps 
an element of $\CS^4_\lambda$ to an element of $C(S^4)$, 
with the identification of every element with $a,a^*$ 
to the left of $b,b^*$ with the corresponding function on 
$S^4$.  Then $x \mapsto \int \eta (x)$ is a trace on 
$\CS^4_\lambda$.
\end{proposition}

Clearly we have a linear map, it remains only to show 
the cyclicity. First, note that the integral on $S^4$ 
is nontrivial on functions depending only on $aa^*$ 
and $bb^*$. Therefore, we might restrict ourselves to 
such case. Let us take two monomials $p,q$ in 
$a,a^*,b,b^*$ such that their product is a monomial 
of $aa^*$ and  $bb^*$. Then we shall prove that 
$\eta(pq) = \eta(qp) $. Let $p= a^{\alpha_p} (a^*)^{\beta_p}
b^{\gamma_p} (b^*)^{\delta_p}$ and $q= a^{\alpha_q} (a^*)^{\beta_q}
b^{\gamma_q} (b^*)^{\delta_q}$.
First, we calculate $pq$ using (\ref{algebra}):
$$ p\,q = \lambda(t)^{\gamma_p \beta_q + \delta_p \alpha_q} 
\bar{\lambda}(t)^{\gamma_p \alpha_q + \delta_p \beta_q} a^{\alpha_p +\alpha_q} 
(a^*)^{\beta_p+\beta_q} b^{\gamma_p+ \gamma_q} (b^*)^{\delta_p+\delta_q},$$
since $\bar{\lambda} = \lambda^{-1}$ we might rewrite the formula as:
$$ \eta(p\, q) = \lambda(t)^{(\gamma_p-\delta_p) (\beta_q - \alpha_q) } 
\eta(p) \eta(q). $$
On the other hand, for $qp$ we have:
$$ q\, p  = \lambda(t)^{\gamma_q \beta_p + \delta_q \alpha_p} 
\bar{\lambda}(t)^{\gamma_q \alpha_p + \delta_q \beta_p} a^{\alpha_q +\alpha_p} 
(a^*)^{\beta_q+\beta_p} b^{\gamma_q+ \gamma_p} (b^*)^{\delta_q+\delta_p},$$
which gives:
$$ \eta(q\, p) = \lambda(t)^{(\gamma_q-\delta_q) (\beta_p - \alpha_p) } 
\eta(p) \eta(q). $$
Now, it is easy to see that both coefficients are equal, since by our
assumption that the product depends only on $aa^*$ and $bb^*$:
$$ \alpha_p + \alpha_q = \beta_p + \beta_q, \;\;\;\;
\gamma_p + \gamma_q = \delta_p + \delta_q,$$
and thus:
$$ (\gamma_p-\delta_p) (\beta_q - \alpha_q) =
(\gamma_q-\delta_q) (\beta_p - \alpha_p).$$

We now define the integral on $4$-forms.
\begin{proposition}
There exist a linear functional on $\Omega^4(\CS^4_\lambda)$ such
that $\int (d\rho) =0$ for every $\rho \in \Omega^3(\CS^4_\lambda)$
and $\int \omega = \frac{8}{3} \pi^2$.
\end{proposition}
\begin{proof}
We begin by defining the integral. Since we know that every 
four-form $\theta$ could be written as $\theta = x \omega$ we 
shall set
\begin{equation}
\int \theta = \int \eta(x).
\end{equation}

Note that since $\omega$ is central, $x\, \omega = \omega\, x$,
we have in effect a linear map $\eta: \Omega^{4}(\CS^4_\lambda) 
\mapsto \Omega^4(S^4)$. We shall demonstrate that there
exists also the extension of map $\eta: \Omega^3(\CS^4_\lambda) \mapsto 
\Omega^3(S^4)$ such that the following diagram is commutative:
$$
\begin{CD}
 \Omega^3(\CS^4_\lambda) @>d>>  \Omega^{4}(\CS^4_\lambda)  \\
 @V\eta VV @VV\eta V \\
  \Omega^3(S^4) @>d>>  \Omega^{4}(S^4)  
\end{CD}
$$

To define the map $\eta$ on three forms we shall use 
their following presentation as a linear space:
\begin{observation}\label{order}
Every $3$-form (over polynomials) could be presented 
(though not in unique way) as a finite sum of elements 
of the type:
$$ \rho = t^\alpha p(a,a^*) \, \chi_i \, q(b,b^*)$$
where $\alpha=0,1$ and $\chi_i$ are forms of 
the type:
\begin{eqnarray*}
da \, da^* \, db, & da \, da^* \, dt, & da \, da^* \, db^*, \\
da \, db \, db^*, & da^* \, db \, db^*, & dt \, db \, db^*, \\
dt \, da \, db, &  & dt \, da \, db^*, \\
dt \, da^* \, db, & & dt \, da^* \, db^* .
\end{eqnarray*}
Of course, these forms are not independent (when we consider
them in the bimodule of three-forms). However, it is important
that we can map them to $\Omega^3(S^4)$ by setting first
$\eta(\chi_i)$, for instance:
$$ \eta ( da\, da^*\, db) = d\eta(a)\, d\eta(a^*)\, d\eta(b),$$
and then:
$$ \eta(\rho) = \eta(t)^\alpha \, p(\eta(a),\eta(a^*)) 
\, \eta(\chi) \, q(\eta(b), \eta(b^*)). $$
To see that the map is well-defined (as a linear map) 
let us observe that by using the so ordered product 
of functions and differentials we  see no
nontrivial commutation rules. Thus, the characterization 
of $\Omega^3(\CS^4_\lambda)$ and $\Omega^3(S^4)$ 
as a linear space are exactly the same. 
\end{observation}

Now, using the presentation (\ref{order}) we can easily see 
that $\eta(d \rho) = d \eta(\rho)$ for every three-form $\rho$. 
Indeed, the external derivative vanishes on all three-forms 
$\chi$ and on functions depending only of $a,a^*,t$ and respectively, 
on $b,b^*,t$ we have standard differentiation:
$$ d \eta( p(a,a^*,t) ) = \eta( dp(a,a^*,t) ), $$
and 
$$ d \eta( q(b,b^*,t) ) = \eta( dq(b,b^*,t) ). $$
Since again, we multiply by $a,a^*$ and its differentials
from the left and $b,b^*$ and its differentials from the 
left -- we encounter no commutators between $a,a^*$ and
their differentials and $b,b^*$ and their differentials. 
Hence, noncommutativity plays no role in the map $\eta$ 
and the action of the external derivative.
\end{proof}

Using the constructed differential structures and the trace 
we have:

\begin{proposition}
$\Omega^*(\CS^4_\lambda)$ is a differential graded 
algebra with a closed graded trace 
$\int : \Omega^4(\CS^4_\lambda) \to \C$.

\end{proposition}
\begin{proof}
So far we have showed the existence of a closed trace on 
$\Omega^4(\CS^3)$. Because of its particular form (\ref{grtrace})
it is evident that  $\int x \rho = \int \rho x$ for every four-form
$\rho$ and $x \in \CS^4_\lambda$. 

Now, let us take a three-form $\beta$ and a one-form $x\, dy$:
\begin{equation}
\begin{array}{l}
\int (x\, dy\, \beta + \beta \, x \, dy) = \int ( dy \, \beta \, x + \beta \,d(xy)
- \beta \, dx \, y)  \\
\phantom{xxx}  = \int ( d( y \, \beta \, x) - y \, d(\beta \, x)
+ \beta\, d(xy) + d(\beta \, x) y - (d\beta) \, xy = \\ 
\phantom{xxx} =  \int \left( d( y \, \beta \, x) - [ y, d(\beta\, x)] +d ( \beta\, xy) \right) = 0.
\end{array}
\end{equation}
\end{proof}

Similarly, we proceed for two-forms. As an immediate corollary we have:
\begin{corollary}
Let $\psi$ be a multilinear functional defined as:
$$\psi(a_0, a_1, a_2, a_3, a_4) = \int a_0 \, da_1 \, da_2 \, da_3 \, da_4.$$
then $\psi$ is a cyclic cocycle.
\end{corollary}

Having a cyclic cocycle enables us to calculate the Chern-Connes 
pairing with the instanton projector, which we introduced 
earlier (\ref{proj}).

\subsection{The Chern character}

Let us consider the construction of an 
element of $\Omega^4(\CS^4_\lambda)$ out of the projector $e$:
$$ \hbox{ch}(e) = - \frac{1}{8 \pi^2} \hbox{Tr} 
(e\, de\, de\, de\, de), $$
where the trace is over matrix indices of $e$.

We shall use the block form of $e$ and the rules of
differential calculi to facilitate the calculations. 
Let us denote:
$$ q = \left( \begin{array}{ll} a & b \\ -\lambda(t) b^* & 
a^* \end{array} \right), $$
then  we can write $e$ and $de$ as block matrices:
$$ e = \oh \left( \begin{array}{ll} t+1 & q \\ q^* & 1-t \end{array} \right), $$
$$ de = \oh \left( \begin{array}{ll} dt & dq \\ dq^* & -dt \end{array}\right). $$
where $1 \pm t$ and $\pm dt$ denote diagonal matrices.
Using this fact and that $(dt)^2=0$ and $dt$ anticommutes with the
rest of the one-forms, we obtain:
$$ de\, de\, de\, de = \frac{1}{16} \left( 
\begin{array}{ll} (dq\, dq^*)^2 & 4\, dt\, dq\, dq^*\, dq \\
-4\, dt \, dq^* \, dq\, dq^* & (dq\, dq^*)^2 \end{array} \right). $$

Therefore for the trace of $e\, ed\, de\, de\, de$ we shall have:
\begin{equation}
 \begin{array}{l}
\ldots =\frac{1}{32} \hbox{Tr} \left( (1+t) (dq\, dq^*)^2 
+ (1-t) (dq^* \, dq)^2 + \right. \\
\phantom{xxxxxxx} - \left. 4 q\, dt\, dq^* \, dq\, dq^* + 4q^* \,
dt\, dq\, dq^* \, dq \right),
\end{array} \label{chern1}
\end{equation}
where the trace is now over two-dimensional matrices. As a next
step let us calculate $dq\, dq^*$ and $dq^* \, dq$:

{\small
$$
\begin{array}{l}
dq \, dq^* = \\
= \left( \begin{array}{ll} 
da\, da^* + db\, db^* & 2 db\, da - \oh \dot{\lambda}
\bar{\lambda}\, dt ( db\, a + \bar{\lambda} da\, b) \\
2 da^*\, db^* - \oh \dot{\lambda} \, dt ( db^* \, a^* + \bar{\lambda}
da^*\, b^*) & da^* \, da + db^*\, db + \dot{\lambda}
\bar{\lambda}\, dt\, (db b^* +  db^* \, b)  \end{array} \right). 
\end{array} $$
}

Now, we shall calculate the diagonal part of $(dq \, dq^*)^2$, the
element from the top-left corner, $\{(dq \, dq^*)^2\}_{11}$. is:
$$ \begin{array}{l}
\{(dq \, dq^*)^2\}_{11} = (da\, da^* + db\, db^*)^2 + 4 db\, da\, da^*\, db^* + \\
\phantom{x}
- \dot{\lambda} \bar{\lambda}\, dt ( -a\, da^* \, db \, db^* 
+ da\, da^* db^* \, b + da\, da^* db\, b^* - a^*\, da\, db\, db^*) = \ldots
\end{array}$$
In the last expression,  using (\ref{diff2}) we can substitute 
$-a\,da^*  - a\, da$ by $ b\, db^* + b^* \, db + 2t\, dt$, then, however,  
we shall encounter at least one element of the type $(dt)^2$, $(db)^2$
or $(db^*)^2$ and therefore it shall vanish. Moreover, using 
the previously derived rules (\ref{3forms}) we see that in the 
end we obtain: 
$$ \ldots = 6 da\, da^* db\, db^*. $$

Quite similarly, for the other diagonal element of 
$(dq\, dq^*)^2$ we shall have:

$$
\begin{array}{l} 
\{(dq \, dq^*)^2\}_{22} = 4  da^*\, db^* \, db \, da 
+ (da^*\, da + db^* \, db)^2 =  \\
\phantom{xxxx} = 6 da\, da^* db\, db^*. \end{array} $$

The calculation for the sum of the diagonal 
elements of $(dq^* \, dq)^2$
yields (we skip the intermediate technical steps, which are same
as in the previous example):
$$ \hbox{Tr} (dq^* \, dq)^2 = - 12 da\, da^* \, db \, db^*. $$

Coming back to our expression (\ref{chern1}) it is easy to
demonstrate that $-4 \hbox{Tr}( q\, dt\, dq^* \, dq\, dq^*)$
and $ 4  \hbox{Tr} (q^*\, dt\, dq \, dq^*\, dq)$ give the same contributions, which together add up to:
$$ 24\, dt \left( -a\, da^*\, db\, db^* - da\, da^* \, db^*\, b 
+ da\, da^* \, db \, b^* + a^* \, da\, db \, db^* \right). $$

Summing it all together and using again (\ref{diff2}) we obtain:
\begin{equation}
\begin{array}{l}
\hbox{ch}(e) = - \frac{1}{8\pi^2} \int \hbox{Tr}(e\, de\, de\, de\,de) = \\
\phantom{xxxx}
- \frac{1}{8\pi^2} \frac{1}{32} 24 \int \left( t \, da \, da^*\, db\, db^* 
- 2 a\, dt\, da^* \, db\, db^* + 2 dt\, da\, da^*\, db b^* \right) = \\
\phantom{xxxx} = - \frac{3}{32\pi^2} 4 \int \omega = 
- \frac{3}{8\pi^2} \frac{8}{3} \pi^2 = -1.
\end{array}
\end{equation}
where we have used the normalization of the integral of $1$
over $S^4$ giving the volume of four-sphere.

As an immediate corollary we have:
\begin{corollary}
The element $e\, de\,de\,de\,de$ gives a nontrivial cohomology
class of the complex $\Omega(\CS^4_\lambda)$.
\end{corollary}

Now, we shall come back to the first Chern form:
$$ \hbox{ch}_1(e) = - \frac{1}{2\pi i} \hbox{Tr} ( e\, de \, de ), $$
which, evidently, does not vanish:
\begin{equation}
 - \frac{1}{2\pi i} \hbox{Tr} ( e\, de \, de ) = - \frac{1}{2\pi i}
2 \dot{\lambda}(t) \bar{\lambda}(t)\, dt ( b \, db^* + b^*\, db),  \label{ch1f}
\end{equation}
however,  it is in the trivial cohomology class. If $\lambda= e^{- i \phi(t)}$ 
for a real function $\phi$ then:
\begin{equation}
\hbox{ch}_1(e) = \frac{1}{\pi} d \left( \phi(t) 
( b \, db^* + b^*\, db) \right).
\end{equation}

What does it mean? Let us remind that the $ch_1(e)$ in the reduced
$(b,B)$ double complex was clearly a cycle. Furthermore, one might
easily observe that it was depending only on the commutative 
subalgebra generated by $t$, $b$ and $b^*$, which we shall 
denote by $\C[b,b^*,t]$ (we might equally well describe the algebra
as the subalgebra of smooth functions on $S^4$ invariant under 
the action of $\delta_2$ - and it is the algebra of smooth functions 
on a three-dimensional closed ball).

Since it is a regular commutative algebra we might use the results 
relating Hochschild and homology of with the de Rham complex.

\begin{proposition}
There exists an element  $\chi \in C_1(\C[b,t])$ and $\xi \in C_3(\C[b,t])$
such that:
\begin{equation}
ch_1(e) = B \chi + b \xi. \label{coh}
\end{equation}
\end{proposition}
\begin{proof}
First, let us observe that since $b \, ch_1(e) = 0$ we might map 
$ch_1(e)$ to $\Omega^2(\C[t,b])$, the image being exactly the 
two-form (\ref{ch1f}). This form is exact, as we have demonstrated 
explicitly. If we take the one form in $\Omega^1(\C[b,t])$, $\chi_0$,
$d \chi_0 = ch_1(e)$, by using the commutative diagram relating
Hochschild homology with differential forms (see Proposition
2.3.4, p.69, \cite{Loday}) we obtain the desired cycle 
$\chi = \pi^{-1}(\chi_0)$.

Then the Hochschild class of $B \chi$ is the same as this 
of $ch_1(e)$, so the difference is in the image of $b$, and then
by chosing any suitable cycle $\xi$ we get (\ref{coh}).
\end{proof}

Therefore, although $ch_1(e)$ does not vanish identically, we still
are almost in the same situation. By correcting slightly $ch_2(e)$
we are again able to obtain a Hochschild cycle of dimension $4$,
which corresponds to the volume form:
\begin{equation}
v = ch_2(e) + B \xi.
\end{equation}

Indeed:
\begin{equation}
\begin{array}{l}
b v = b \, ch_2(e) + b B \, \xi  = B ch_1(e) - B b \xi \\
\phantom{xxxx} = B(ch_1(e) - b \xi) = B (B \chi) = 0.
\end{array}
\end{equation}

\section{Conclusions}

The construction presented in this paper extends the notion 
of noncommutative spheres to objects defined through 
instanton bundles, whose first Chern class does not vanish
but is homologically trivial. Our aim was to demonstrate that
such solutions exists, are easily obtained by a slight 
generalization of the {\em twisted} noncommutative spheres.
We demonstrated as well the existence of 4-dimensional 
differential calculus (a 4-dimensional cycle) and calculated
explicitly the Chern-Connes pairing.

Of course, it is possible to consider further generalizations
going in this direction, for instance one might consider 
(in the same spirit) the Matsumoto \cite{Matsumoto} 3-spheres 
defined through generators as:

\begin{equation}
\begin{array}{lll}
{}[a ,a^\ast] =0, & \phantom{xxxxx} & [b, b^\ast] =0, \\
ab = \lambda ba, &  & a b^\ast = \bar{\lambda} b^\ast a, \\
a^\ast b = \bar{\lambda} a^\ast b, & & a^\ast b^\ast =
\lambda b^\ast a^\ast,
\end{array} 
\end{equation}
and
\begin{equation}
a a^\ast + b b^\ast = 1, 
\end{equation}
where $\lambda(t)$ is a uni\-tary ele\-ment from the 
center of the algebra, $\lambda(t) \bar{\lambda}(t) =1$, 
for instance:
$$ \lambda = \lambda (bb^*).$$

Similarly as for the four-sphere one may view this algebra as
generated by the matrix elements of is generator of $K_1$
class:
$$
U = \left( \begin{array}{ll} a & b \\ - \lambda b^* & a^* \end{array} \right),
$$
Now, it is easy to verify that the Chern character of the generator $U$
for this algebra is:

$$ ch_{\frac{1}{2}}(U) = b \ts b^* - b^* \ts b
 + \lambda b^* \ts \bar{\lambda} b -  \bar{\lambda} b \ts \lambda b^*. $$
 
Again, although this Chern character does not vanish, since it is over 
a commutative subalgebra we see that the same argument as in the 
case of 4-sphere applies and it is sufficient to study the image 
of $ch_{\frac{1}{2}}(U)$ in the de Rham complex:

$$ \pi(ch_{\frac{1}{2}}(U)) = - \frac{1}{2 \pi i} 
bb^*( \lambda\, d \bar{\lambda} - \bar{\lambda} \, d \lambda).$$

If $\lambda = e^{2 \pi i f(bb^*)}$ for some smooth real 
function $f$ we get:

$$ \pi(ch_{\frac{1}{2}}(U)) = - 2 bb^* f'(bb^*) d (bb^*). $$

To proceed further we need to identify the commutative algebra we
are working with and it is easy to see that these are functions on 
a disk. For this reason the above one-form, which is closed is also
exact - so again, within the de Rham complex the lower Chern 
character is of trivial cohomology class.

Although we have concentrated in this paper only on the case of
four-dimensional spheres (motivated by the instanton algebra 
construction of \cite{CoLa}) there are numerous examples of 
other deformation of this type (one of which we already 
mentioned). Clearly, the procedure might be as well generalized
to higher-dimensional spheres. 

Their applications to physical theories (allowing, for instance, 
for a change of commutativity with time) shall be discussed 
elsewhere \cite{Sit2}. 

{\bf Acknowledgements} \\
The author would like to thank Michel Dubois-Violette for 
discussion and remarks, Piotr Hajac for thorough discussions 
on Matsumoto spheres, H-J.Schneider and J.Wess for kind 
invitation to their seminars and the entire Munich group 
(Lehrstuhl J.Wess) for hospitality.

\vspace{1cm}

\null

\end{document}